# Automatic Modulation Recognition of PSK Signals Using Nonuniform Compressive Samples Based on High Order Statistics

Zhengli Xing[a], Jie Zhou[a*], Jiangfeng Ye[a], Jun Yan[a], Lin Zou[b], Qun Wan[b]
[a]Institute of Electronic Engineering, China Academy of Engineering Physics, Mianyang, China
[b]Department of Electronic Engineering, University of Electronic Science and Technology of China, Chengdu, China
Email: author_correspond@163.com

*Abstract*—The $N^{th}$ Power Nonlinear Transformation (NPT) is a common method for automatic modulation classification, especially for PSK signals. However, greater than Nyquist rate sampling is essential for features extraction in NPT. In this paper, introducing the compressive sensing (CS) theory, we propose a novel Automatic Modulation Recognition (AMR) method based on the frequency spectrum of the $N^{th}$ power nonlinear transformation of PSK type signals, from nonuniform compressive samples. Here, analysis of frequency spectrum reconstruction of the $N^{th}$ power nonlinear transformation of PSK signals is presented, which can be further used for AMR and rough estimations of unknown carrier frequency and symbol rate.

I. Introduction

Automatic Modulation Recognition (AMR) is a fundamental element of cognitive radio [1], which distinguishes the modulation types of received signals. It has been widely used for spectral monitoring and user identification in spectrum sensing. As a hotspot, various techniques are proposed for AMR [1]-[3], such as Wavelet/ Fourier transform, Cumulants. However, sampling rate in most schemes requires higher than Nyquist rate, which causes tremendous burden for ADC, especially for wideband signals. Furthermore, the aims of AMR and carrier frequency and symbol rate estimation only need to extract little information from the enormous data.

Recently, Compressed Sensing (CS) has been introduced as a new sampling theory [4]-[6], which can just extract interested information and recover the wideband signals with a small amount of samples. In particular, for signals that can be sparse-represented in a certain domain, the number of required measurements can be very small.

In this paper, we propose a new method for AMR with Nonuniform Compressive Samples (NCS). Currently, most of the CS applications focus on sampling number reduction, while little attention has been paid to AMR based on CS. Chia and Wakin [7] tried to estimate the $n^{th}$ power spectrum without reconstruction. However, on the basis of CSP theory proposed by Davenport, Wakin, et al [8], this method does not have strange anti-noise ability – it can only exploit the peaks of carrier frequency at high SNR to discriminate BPSK, QPSK, 8PSK signals, and just estimate the carrier frequency. Similarly, combining CS with other traditional methods, Chai raised a new method in [9] to calculate the compressive higher order cyclostationary statistics; based on Tian's work [10], Zhou and others derived some methods [11]-[12] for AMR by reconstructing cyclic spectrum from NCS.

Traditionally, NPT is a classical method in AMR for PSK, ASK, MSK. Nevertheless, the sampling rate required for NPT is several times of Nyquist rate. Based on the previous work [3], [7], in this work, we analyze the $2^{nd}$, $4^{th}$ and $8^{th}$ power spectrum of PSK type signals, including BPSK, QPSK, 8PSK, OQPSK, and MSK, to find different features of peaks for such signals, and we also establish the relationship between NPT and NCS. Thus, we can reconstruct the nonlinear transformed spectrum of PSK type signals to fulfill AMR.

The remainder of this paper is organized as follows. Section II presents the analysis of the $2^{nd}$, $4^{th}$ and $8^{th}$ nonlinear transformations of PSK type signals with uniform sampling in the time domain. It describes the relationship between NCS and the nonlinear transformation and spectrum reconstructions in Section III. Section IV gives the classifier and rough estimation methods of carrier frequency and symbol rate. Simulation results are presented in Section V. Finally, conclusions are made in Section VI.

II. Features of NPT For PSK Type Signals

Detecting the number of peaks in the spectrum after nonlinear transformation is the main task of NPT method for AMR. With the number of peaks, we can distinguish different kinds of signals. Besides, by seeking the locations of the peaks, we can also roughly estimate the carrier frequency and symbol rate.

Here, we extend the analysis to 8PSK and OQPSK from [3] and summarize the spectrum features of NPT signals for PSK modulation.

A. The Signal Model

In this section, the general model of PSK type signals is defined.

For MPSK, the signal model is as follows [3], [11]:

$$s_{\text{MPSK}}(t) = \sum_{n=-\infty}^{\infty} A g(t-nT_s) \exp(j2\pi \frac{m_n-1}{M} + j2\pi f_c) \quad (1)$$

where $A$ is the amplitude; $T_s = 1/R_s$ is the symbol period, and $R_s$ is the symbol rate. $M \in \{2,4,8\}$ is the number of unique phases; $m_n$ is the $n^{th}$ transmitted symbol; $f_c$ is the carrier frequency; g(t) is the Root Raised Cosine Pulse Shape (RRC); $\alpha$ is the roll-off factor.

For OQPSK, the signal model is:
$$s_{OQPSK}(t) = A[I_n + j*Q_n]\exp(j2\pi f_c t) \quad (2)$$

where
$$I_n = \sum_n a_n g(t-(2n-1)T_b), \quad Q_n = \sum_n b_n g(t-2nT_b) \quad (3)$$

and $a_n, b_n \in \{1,-1\}$ are i.i.d. (independently identically distributed) random sequences; $T_b = 1/R_b = 1/2T_s$, where $R_b$ is the bitrate.

For MSK, the signal model is:
$$s_{MSK}(t) = A[I_n + j*Q_n]\exp(j2\pi f_c t) \quad (4)$$

where
$$I_n = \sum_n a_n \text{rect}(t-(2n-1)T_b)\cos(\frac{\pi t}{2T_b})$$
$$Q_n = \sum_n b_n \text{rect}(t-2nT_b)\sin(\frac{\pi t}{2T_b}) \quad (5)$$

and $\text{rect}(\bullet)$ denotes the rectangle function.

*B. Spectrum Features of NPT Signals*

More details about spectrum features after NPT for BPSK, QPSK and MSK can be found in [3]. With the calculation method in [3], we can also easily infer the spectrum features after NPT for 8PSK and OQPSK, which are shown in Table 1 and 2. From these two tables, different modulation types have different numbers of peaks of spectrum after diverse NPTs. Thus, we can use these features for AMR, as well as carrier frequency and symbol rate rough estimations (except for 8PSK).

### III. FEATURES FROM NCS

*A. CS Sampling*

Low-rate random sampling rather than high-rate uniform sampling, is one significant advantage in CS. Just like previous studies [7], [9]-[11], we can design the measurement matrix of CS theory in our application as a random binary sampling matrix, which contains a single 1 on each row.

Just as shown in Table 1 and Table 2, PSK signals raised to the 2nd and 4th powers are sparse in spectrum domain. So we can choose DFT synthesis matrix as the sparsifying matrix, and the procedure can be written as:
$$\mathbf{y} = \mathbf{\Phi z} = \mathbf{\Phi \Psi f} \quad (6)$$

where $\mathbf{\Phi} \in \mathbb{R}^{M \times L}$ is the measurement matrix. $\mathbf{\Psi}$ is the $L \times L$ DFT synthesis matrix. $\mathbf{z}$ is the length-L vector of uniform samples of the interested analog signal (obeying Nyquist Sampling Theorem), and $\mathbf{y}$ is the length-M vector of nonuniform compressive samples. It has been proved [14]-[15] that, the matrix $\mathbf{A} = \mathbf{\Phi \Psi}$ satisfies the RIP of $s = O(M/\log(L)^4)$.

As discussed above, the DFT coefficients of signals undergone NPT are sparse, thus, we can perfectly recover the corresponding spectrum.

*B. The Relationship Between NPT and NCS*

Here, we define $\mathbf{y}_N$ and $\mathbf{z}_N$ as the NPT forms of $\mathbf{y}$ and $\mathbf{z}$ respectively:
$$\mathbf{y}_N = [y_1^N, y_2^N, \cdots, y_M^N]^T, \quad \mathbf{z}_N = [z_1^N, z_2^N, \cdots, z_M^N]^T$$

Considering the structural property of $\mathbf{\Phi}$. The NPT forms can be rewritten as [11]:
$$\mathbf{y}_N = \mathbf{\Phi z}_N = \mathbf{\Phi \Psi f}_N \quad (7)$$

where $\mathbf{f}_N$ is the DFT coefficients of $\mathbf{z}_N$. Just as discussed above, $\mathbf{f}_N$ is sparse for certain orders of N, when apply NPT to PSK signals. Without directly sampling the sequences of $\mathbf{z}_N$, (7) indicates the relationship between NPT and NCS.

*C. Spectrum Recostruction*

The problem of recovering $\mathbf{f}_N$ from $\mathbf{y}$ or $\mathbf{y}_N$ is a NP-hard problem. It can be converted to $l_1$-norm optimization problem in CS [16]:
$$\hat{\mathbf{f}}_N = \arg\min \|\mathbf{f}_N\|_1 \quad s.t. \quad \mathbf{y}_N = \mathbf{\Phi \Psi f}_N \quad (8)$$

For example, for N=2, the model can be rewritten as:
$$\hat{\mathbf{f}}_2 = \arg\min \|\mathbf{f}_2\|_1 \quad s.t. \quad [y_1^2 \cdots y_M^2]^T = \mathbf{y}_2 = \mathbf{\Phi \Psi f}_2 \quad (9)$$

We can solve the convex problems with CVX [17].

The spectrum of a BPSK signal raised to the power of 2 is shown in Fig. 1 (a). Fig. 1 (b) is the reconstruction result with CVX. In Fig. 1 (a), there are 3 main peaks that are in accordance to the theoretical results given in Tables 1 and 2; as shown in Fig. 1 (b), the three peaks can be completely recovered with CVX.

### IV. CLASSIFICATION AND ESTIMATION

After recovering spectrum of NPT, AMR and rough estimations of $f_c$ and $R_s$ can be fulfilled.

*A. AMR Strategy*

Here, let $c_1$, $c_2$ and $c_3$ denote the numbers of discrete dominant peaks of $\mathbf{f}_2$, $\mathbf{f}_4$ and $\mathbf{f}_8$. According to Table 2, our procedures of AMR technique is revealed in Fig. 2.

*B. Rough Estimation methods of $f_c$ and $R_s$*

As can be seen in Table 1, for each kind of signal, the locations of discrete peaks of $\mathbf{f}_2$ and $\mathbf{f}_4$ are determined by $f_c$ and $R_s$ except 8PSK. The locations of discrete peaks can be used to roughly estimate $f_c$ and $R_s$. Since different kinds of signals correspond to different locations, here, for instance, we give the estimates of $f_c$ and $R_s$ for QPSK. Other kinds of signals can be obtained in a similar way.

Here, let $A_1, A_2, A_3$ denote the locations of three dominant peaks of $\mathbf{f}_4$ of the QPSK signal. It's obvious that,

Table 1. Peak lines in different nonlinearities for PSK type signals
N: None. Y: Exist ($n \in \mathbb{Z}$)

| Nonlinearity | Frequency | Modulation Type | | | | |
|---|---|---|---|---|---|---|
| | | BPSK | QPSK | 8PSK | OQPSK | MSK |
| None | $f_c$ | N | N | N | N | N |
| | $f_c + nR_s$ | N | N | N | N | N |
| $(.)^2$ | $f_c$ | Y | N | N | N | N |
| | $f_c + (n+0.5)R_s$ | N | N | N | N | Y |
| | $f_c + nR_s$ | Y | N | N | Y | N |
| $(.)^4$ | $f_c$ | Y | Y | N | Y | N |
| | $f_c + (n+0.5)R_s$ | N | N | N | N | N |
| | $f_c + nR_s$ | Y | Y | N | Y | Y |

Table 2. Number of discrete peaks for PSK type signals

| Nonlinearity | Modulation Type | | | | |
|---|---|---|---|---|---|
| | BPSK | QPSK | 8PSK | OQPSK | MSK |
| None | 0 | 0 | 0 | 0 | 0 |
| $(.)^2$ | 3 | 0 | 0 | 2 | 2 |
| $(.)^4$ | 3 (or 5) | 3 (or 5) | 0 | 3 (or 5) | 2 |
| $(.)^8$ | 3 (or 5) | 3 (or 5) | 3 (or 5) | - | - |

$$\hat{f}_c = A_1 = (A_2 + A_3)/2$$
$$\hat{R}_s = |A_1 - A_2|/2 \quad (10)$$

## V. SIMULATION RESULTS

The proposed methods are tested and verified in this section. Simulation scenarios are set as follows: symbol number is 1024, $\alpha = 0.5$, $f_c = 500Hz$, $R_s = 800Hz$. Here, the Nyquist-rate of uniform sampling is $f_s = 6.4kHz$ to avoid aliasing of spectrum after NPT, which means $L = 8192$. And the uniform samples correspond to the "Nyquist rate" curves in Fig. 3. Besides, the number of NCS is $M = 0.3 \times L = 2458$.

Fig. 3 (a) and Fig. (b) depict the rate of correct classifications with varying signal-to-noise ratios (SNR). Fig. 3 (c) shows the estimation results of the carrier frequency of QPSK signal. Here, we can see that, for a given probability of correct classification, AMR using NCS requires another 2-6 dB of SNR, compared with uniform sampling. And for the estimation accuracy of $f_c$, the requirement is about 5dB.

## VI. CONCLUSION

Inspired by the CS theory, we combine NCS with NPT method to solve the problems of AMR and estimations of carrier frequency and symbol rate. Simulation results show that the method we proposed can efficiently complete the tasks.

## VII. ACKNOWLEDGEMENTS

This work is supported in part by the NSFC (Grant No. 61301267), and in part by the Fundamental Research Funds for the Central Universities.

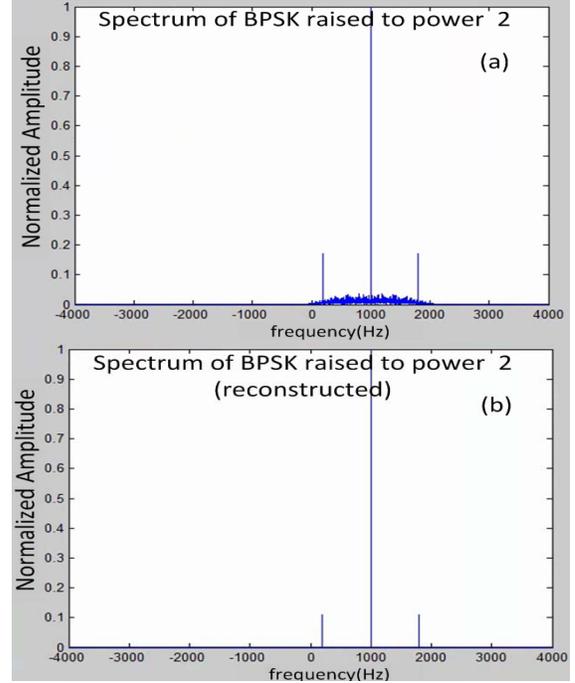

Fig. 1. Simulation results for BPSK signals after NPT

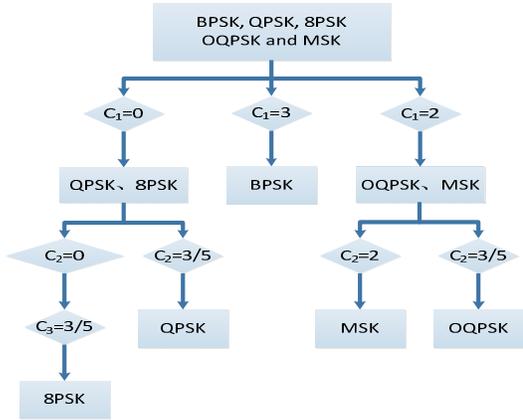

Fig. 2. Structure of Classifier for AMR

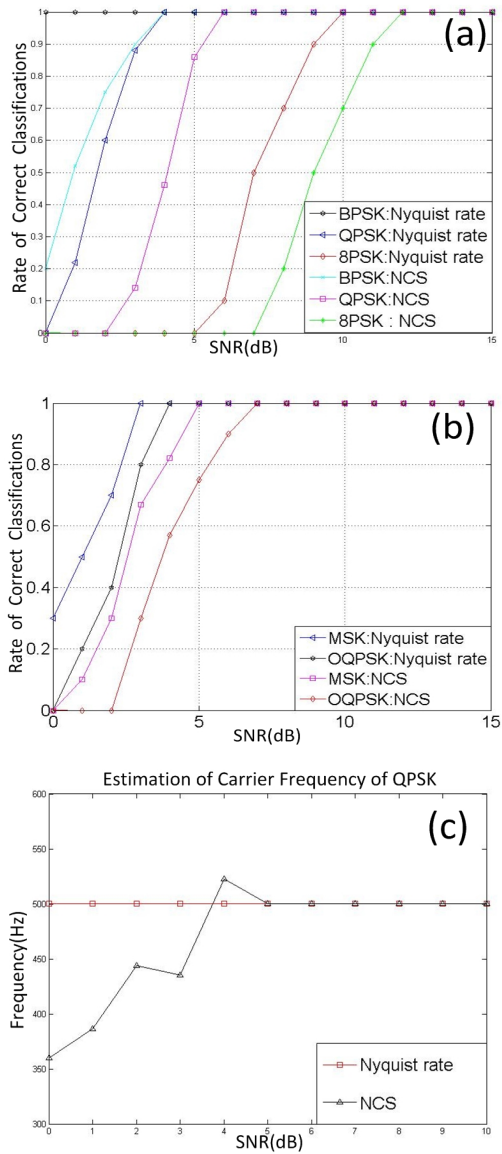

Fig. 3. Simulation results of AMR and estimation of $f_c$